\documentstyle[12pt]{article}
\def\gsim{{\ \lower 0.6ex \hbox{$\buildrel>\over\sim$}\ }}
\begin{document}

\begin{titlepage}
\vspace*{-2cm}
\begin{flushright}
FERMILAB-Pub-93/086-T\\
CERN-TH-6868/93\\
DTP/93/24\\
hep-ph/9305220\\
\end{flushright}
\vspace{1cm}
\begin{center}
{\Large\bf Single Lepton Production at Large Transverse Momentum}\\
\vspace{1cm}
{\large
W.~T.~Giele}\\
\vspace{0.5cm}
{\it
Fermi National Accelerator Laboratory, P.~O.~Box 500,\\
Batavia, IL 60510, U.S.A.} \\
\vspace{1cm}
{\large
E.~W.~N.~ Glover}\\
\vspace{0.5cm}
{\it
Physics Department, University of Durham,\\ Durham DH1~3LE, England} \\
\vspace{0.5cm}
and \\
\vspace{0.5cm}
{\large
David~A.~Kosower\footnote{On leave from the Centre d'Etudes
of Saclay, F-91191 Gif-sur-Yvette cedex, France}}\\
\vspace{0.5cm}
{\it
Theory Division, CERN,\\ CH--1211 Geneva 23, Switzerland  \\
\tt kosower@nxth02.cern.ch}\\
\vspace{0.5cm}
{\large April 1993}
\vspace{0.5cm}
\end{center}
\begin{abstract}
We study the production of single isolated
leptons at large transverse momentum, $p_T^\ell > M_W/2$.
The dominant source of such leptons is production of an on-shell $W$ boson
recoiling against a hard jet.
Vetoing this jet forces the $W$ boson to be produced off
resonance and significantly reduces the standard model cross
section, thereby enhancing the discovery prospects for
non standard model physics. A significant number
of events  have the topologically interesting signature of a large $p_T$
lepton  balancing a jet with little measured missing transverse energy.
\end{abstract}

\end{titlepage}

Just over a decade ago, the $W$ and $Z$ bosons  were discovered at the
CERN S$p\bar  p$S by observing  isolated  leptons  at large transverse
momentum.  In the case of the $W$ boson, the lepton was accompanied by
significant missing transverse  momentum due to  the neutrino escaping
the detector.   The cross  section  for these processes  is  roughly a
nanobarn.   In the near  future, experiments  at the Fermilab Tevatron
expect to collect  an integrated luminosity  of  100 pb$^{-1}$ and are
thus  sensitive  to  cross sections of  the order of  0.1~pb.  At such
cross sections,  one  might hope to find  hints  of physics beyond the
standard model.

Let us consider, for example,  events with isolated  leptons and jets,
where the lepton has a large transverse momentum, $p_T^\ell > 70$~GeV,
or events  with leptons  and  missing   transverse energy   where  the
transverse  mass, $M_T(\ell,\nu)   >   140$~GeV.   These  events   are
topologically  interesting  and  could  represent the  signal   of new
physics.  One example is a  $W^\prime$ boson of extended gauge models
\cite{wprime}.
However,
one must first  understand  the   possible sources of   such events in
standard model processes, and that is our purpose in this Letter.

We begin by considering the lowest order process,
\begin{equation}
q \bar q \to W^\pm \to \ell^\pm \nu.
\end{equation}
When  $p_T^\ell \sim  M_W/2$,   the   $W$   boson  can  be    produced
on-resonance.   However, if   we demand  that the  lepton   transverse
momentum be  much  larger,   the  $W$ boson  will  be forced   off its
resonance and the cross  section will fall  sharply.  This is shown in
Fig.~1 where we plot the total cross section for  the  production of a
single species of charged  lepton
with $p_T^\ell  >    p_T^{\rm  cut}$ and   rapidity
$|\eta^\ell |  <  2$ in association with a  missing  transverse energy
$E_T^{\rm miss} > 20$~GeV.
Motivated by the recent though preliminary H1  data on
deep inelastic scattering \cite{de Roeck} we have  restricted our
attention  to  the    MRSD-$^\prime$ \cite{MRS}  structure   functions.
However, since we are dealing with a process at relatively  large $x$,
$x \sim 0.05$, different choices for  the input structure functions do
not make a  significant difference.  We  observe  that for $ p_T^{\rm
cut} = 70$~GeV, the cross section is still about 1~pb.
\begin{figure}[t]\vspace{11cm}
\includegraphics{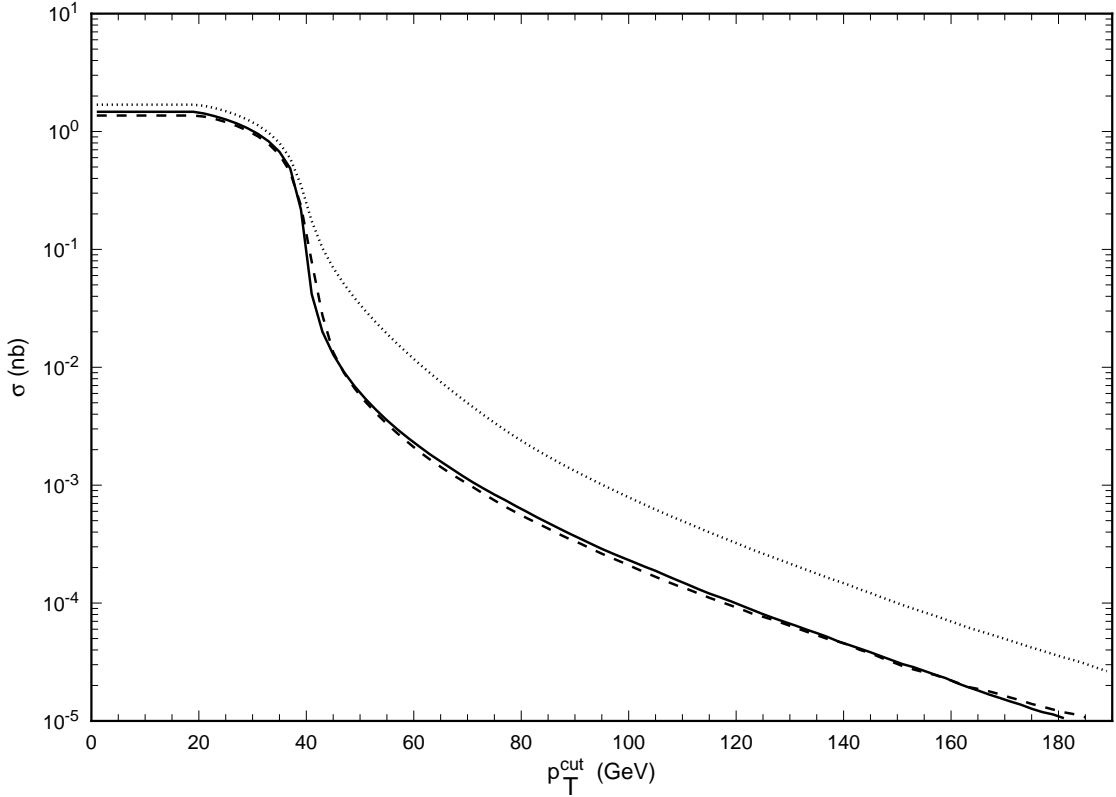}
\caption[]{The total cross section for events containing
leptons with transverse momentum  $p_T^\ell >  p_T^{\rm cut}$
in association with missing transverse energy.
Both leading order (solid line) and next-to-leading order
(dotted line) predictions are shown.   We also show the next-to-leading order
prediction for the jet-veto rate   where no hadronic
cluster with $E_T >  10$~GeV  and  $|\eta  | < 4$ is observed in the event
(dashed line).}
\end{figure}

At next-to-leading order,
one must consider radiative processes such as,
\begin{equation}
q\bar q \to W^\pm g \to \ell^\pm \nu g, ~~~~q g \to W^\pm q \to \ell^\pm \nu q,
\end{equation}
which can  give  rise to a    jet.  In this  case,  the  definition of
$E_T^{\rm miss}$ is slightly modified  to be the transverse imbalance of
all energy deposited in a calorimeter which extends to  $|\eta| = 4$.
Using the results of ref.~\cite{GGK}, we show the next-to-leading order
corrections to the total cross section as a  function of $ p_T^{\rm cut}$
in Fig.~1.  These next-to-leading corrections alter the predicted rate
for the inclusive  process given a large $p_T^{\rm  cut}$, by a  large
factor ---  a factor of nearly five at  $p_T^{\rm  cut}= 70$~GeV.  The reason
for  this increase  is   straightforward; requiring  a   large  lepton
transverse momentum no longer forces the $W$ far  off resonance,
since it
can acquire  a  large transverse momentum  by recoiling against a jet.
This additional transverse momentum  is then inherited by the  daughter lepton.
Thus, there is a trade  off between  a larger subprocess center of mass
energy combined with a factor of $\alpha_s$ and the off-resonance suppression.
To show that this is indeed the
case,  we also show  the cross section  for  the production  of  large
transverse momentum leptons at next-to-leading order where we veto events by
requiring that no hadronic
cluster with transverse energy $E_T >  10$~GeV  and  $|\eta  | < 4$  be
observed. As
expected, the  genuine ${\cal O}(\alpha_s)$   corrections to the 0-jet
rate are  small.  This  gives us confidence in  the convergence of the
perturbation series for this quantity.  We see that the  majority of  events
containing   a large transverse momentum electron   fail this veto.

It is also instructive to examine the transverse mass distribution for
these   events.
The transverse mass $M_T$ is defined as
\begin{equation}
M_T = \sqrt{2p_T^\ell E_T^{\rm miss} \left(1-\cos \phi\right)}
\end{equation}
where $\phi$ is the azimuthal angle between the
lepton momentum and the missing transverse energy vector.
In Fig.~2,    we  show  the    next-to-leading order
predictions for this distribution for both  the inclusive reaction and
reactions with a  jet veto (using  the  same cuts  as in the  previous
paragraph)   and $p_T^\ell >  p_T^{\rm cut}  =  70$~GeV.  The peak at
$M_T\sim M_W$ arises  from  events where  the large lepton   $p_T$  is
balanced by  a  hard  jet.  The second  peak  is   perhaps unexpected,
however.
Applying a
jet  veto eliminates the  first  peak but  not the second,
indicating that the second peak
 arises   when the  transverse momentum of the lepton  and  neutrino
approximately
balance    and  there is  little hadronic   energy in  the
event.
The $W$ boson   is  far off   shell and  this  contribution is
restricted to the region $M_T \gsim 2 p_T^{\rm cut} = 140$~GeV.
The total cross section in the region $M_T > 140$~GeV is  about 1~pb which
should be observable using the current data sample.  But one must
exercise caution in isolating events in this region: the inclusive
spectrum might suggest that the second peak is a signature of new
physics in spite of its artifactual origin in the cuts.

\begin{figure}[t]\vspace{11cm}
\includegraphics{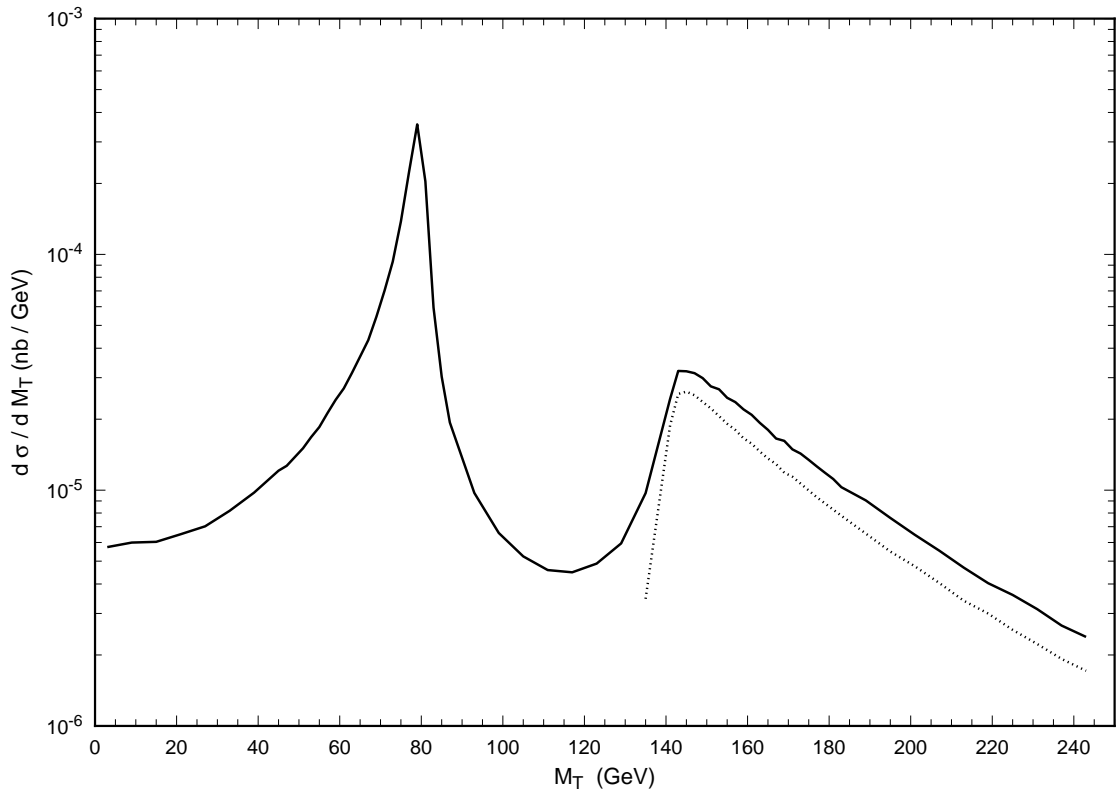}
\caption[]{The inclusive next-to-leading order transverse mass distribution
for events containing a charged lepton with $p_T^\ell > 70$~GeV,
$|\eta^\ell |  <  2$ and
$E_T^{\rm miss} > 20$~GeV.
The rate for events passing a jet veto is shown dotted. }
\end{figure}

One may also consider, as mentioned  in the introduction, the class of
events where the charged lepton recoils against a jet, yet the missing
transverse energy is small.    In Fig.~3, we
show the leading and next-to-leading order predictions for the missing
transverse energy distribution in events where one jet\footnote{The
jet axis and energy are obtained by adding the four-momenta of
partons contained in a cone of size $\Delta R$.}  with
transverse momentum $E_T^{\rm
jet}>   25$~GeV, pseudorapidity $|\eta^{\rm jet}|< 3$
and cone size $\Delta R = 0.7$
is observed in
association   with a charged  lepton
possessing $p_T^\ell  > p_T^{\rm   cut} =   70$~GeV.
The next-to-leading order    QCD
corrections  do  not alter  the shape of the distribution.
  The largest  cross  section occurs when the $W$
boson is on shell. In this case, for  back-to-back lepton and neutrino,
we expect
\begin{equation}
E_T^{\rm miss} \sim \frac{M_W^2}{4 p_T^{\rm cut}} \sim 23~{\rm GeV},
\end{equation}
and a peak near this value can be seen.  Although most events do contain
noticeable  missing tranverse  energy, the   total cross section for
lepton +  jet   events with less  than 20~GeV   of missing energy   is
appreciable (about 2~pb).

\begin{figure}[t]\vspace{11cm}
\includegraphics{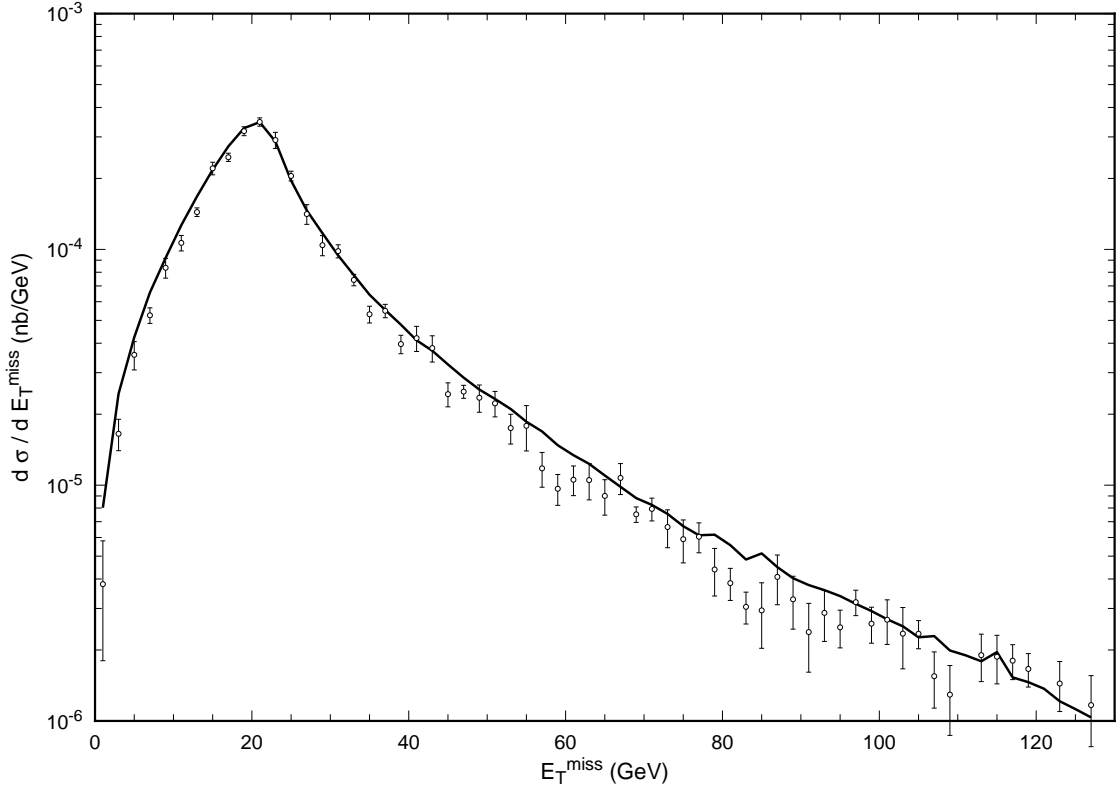}
\caption[]{The leading order $E_T^{\rm miss}$ distribution
for events containing a charged lepton with $p_T^\ell > 70$~GeV,
$|\eta^\ell |  <  2$ and
a single jet $E_T^{\rm jet} > 25$~GeV.
The next-to-leading order rate is shown as points. }
\end{figure}

In this  letter  we have  examined  two classes  of events   with high
transverse momentum leptons, isolated lepton and lepton +  jet events.
One must take care, when imposing a high transverse momentum cut
on the lepton,  to use next-to-leading  order calculations in order to
verify that  deviations from predictions  are indeed a signal  of  new
physics.
We note that the standard model contains other possible backgrounds such as
 $\tau^+\tau^-$ and
heavy flavour production which we have not estimated.

The first class contains no jet, a possible signal for  an object such as
a $W^\prime$ boson  or, alternatively, for deviations  from the  standard model
due to
other sources of  higher-dimension  operators.  In these events, it is
essential to impose a jet  veto.    Such a veto   will not reduce  the
$W^\prime$ boson signal but  will reduce the standard model background
by a large  factor, perhaps as large  as 5.  One  can simply count the
 number  of  lepton events and compare  them with the predictions  of
Fig.~1.  This  cross section  is   largely unaffected  by
next-to-leading   order
corrections; the data with a  low  lepton transverse momentum cut imposed can
be used to normalize the theoretical prediction to the data.

The second class of events contains  a jet in addition to a lepton.
Some exhibit the
striking feature of a high momentum jet balancing a  lepton, while the
missing transverse energy is small.   With a luminosity as small  as
25  pb$^{-1}$ we  can  expect  of  the order   of 50  events with  the
following characteristics:  $p_T^\ell  > 70$~GeV  and $E_T^{\rm   miss} <
20$~GeV balanced by a large $E_T $ jet.

We thank the Fermilab ACPMAPS project for providing computer time
for some of the calculations described above.  In addition,
E.W.N.G.~and D.A.K.~wish to thank the Fermilab theory group for its
hospitality during the completion of the work described above.

\end{document}